\documentclass[twocolumn]{aastex63}

\newcommand{\feh}{\mbox{$\rm [Fe/H]$}}
\newcommand{\logg}{\mbox{$\log g$}}
\newcommand{\teff}{\mbox{$T_{\rm eff}$}}

\received{xx}
\revised{xx}
\accepted{xx}
\submitjournal{AJ}

\shorttitle{Sample article}
\shortauthors{de Mijolla}
\graphicspath{{./}{figures/}}
\usepackage{amsmath}
\DeclareMathOperator{\argmin}{argmin} 
\usepackage{graphicx}
\usepackage{todonotes}
\usepackage{comment}

\begin{document}

\title{Disentangled Representation Learning for Astronomical Chemical Tagging}

\correspondingauthor{Damien de Mijolla}
\email{ucapdde@ucl.ac.uk}

\author{Damien de Mijolla}
\affiliation{Department of Physics and Astronomy,\\
University College London,
\\ Gower Street, WC1E 6BT, UK}

\author{Melissa Ness}
\affiliation{Department of Astronomy,\\ 
Columbia University, \\
Pupin Physics Laboratories, New York, NY 10027, USA}
\affiliation{Center for Computational Astrophysics, Flatiron Institute, 162 Fifth Avenue, New York, NY, 10010, USA}

\author{Serena Viti}
\affiliation{Leiden Observatory,\\ Leiden University,\\ PO Box 9513, 2300 RA Leiden, Netherlands}
\affiliation{Department of Physics and Astronomy,\\
University College London,
\\ Gower Street, WC1E 6BT, UK}

\author{Adam Wheeler}
\affiliation{Department of Astronomy,\\ 
Columbia University, \\
Pupin Physics Laboratories, New York, NY 10027, USA}



\begin{abstract}
Modern astronomical surveys are observing spectral data  for millions of stars. These spectra contain chemical information that can be used to trace the Galaxy's formation and chemical enrichment history. However, extracting the  information from spectra, and making precise and accurate chemical abundance measurements are challenging. Here, we present a data-driven method for isolating the chemical factors of variation in stellar spectra from those of other parameters (i.e. \teff, \logg, \feh). This enables us to build a spectral projection for each star with these parameters removed. We do this with no ab initio knowledge of elemental abundances themselves, and hence bypass the uncertainties and systematics associated with modeling that rely on synthetic stellar spectra. To remove known non-chemical factors of variation, we develop and implement a neural network architecture that learns a disentangled spectral representation. We simulate our recovery of chemically identical stars using the disentangled spectra in a synthetic APOGEE-like dataset. We show that this recovery declines as a function of the signal to noise ratio, but that our neural network architecture outperforms simpler modeling choices. Our work demonstrates the feasibility of data-driven abundance-free chemical tagging.
\end{abstract}



\section{Introduction} \label{sec:intro}

Galactic archaeology, the sub-field of astronomy interested in reconstructing the Galaxy's history, has recently experienced substantial growth. This has been spurred by stellar surveys such as RAVE, APOGEE, GALAH, LAMOST, GAIA and GAIA-ESO \citep{Rave2006, Apogee,Galah,Lamost,Gaia,GaiaESO1,GaiaESO2}. These surveys have obtained spectra, and in the case of Gaia,  astrometry and photometry, for hundreds of thousands to millions of stars across the Galaxy. These data have enabled measurement of stellar abundances, distances and ages across the Galaxy. Future missions are also on the horizon \citep{4most,SloanV,Weave,pfs}. 

Chemical element abundances derived from stellar spectra are core to archaeological pursuits. While there are evolutionary and environmental factors which can impact the surface abundance of a star \citep[e.g.][]{AtomicDiffusionLiu, Casey2019}, abundances link stars to individual molecular clouds, which give their stellar brood with similar chemical fingerprints \citep{2014Natur.513..523F, Bovy2016, Krumholz2019,  Ness2018, AtomicDiffusionLiu}. The chemical space of stars in the Milky Way's disk seems fairly low-dimensional, with stars born at the same radius and time  being chemically similar or even identical within measurement precision \citep{Ness2019, Weinberg2019, Ting2012, Price-Jones2019}. Indeed, at solar metallicity, the APOGEE survey shows that 1 percent of field stars are as chemically similar as stars that are known to be from the same individual birth cluster \citep{Ness2018}. This doppelganger rate alone renders chemical tagging of stars to their individual birth sites, using $\approx$ 20 abundances alone, rather difficult. 
Nevertheless, identifying chemically identical or near-identical stars, has high utility in reconstructing the galaxy's formation. For example, in estimating the number of star-forming clusters in the galactic disk (e.g. \cite{Kamdar} and \cite{Ting2016}) or for understanding how stars have moved over time (e.g. \cite{Beane, Coronado2020, Price-Jones2020} and \cite{2018ApJ...865...96F}). Furthermore, detailed abundances allow connecting stars to their birth radii as well as their time of formation \citep{Ness2019, Bedell2018, Feuillet2019, Casali2020}
Typically, efforts to identify chemically identical stars have involved estimating surface abundances by comparing observations to synthetic spectra, and then running a clustering algorithm \citep{Price-Jones2019,Hogg2016}. This procedure is hampered by its reliance on imperfect stellar models to obtain the abundance labels that describe the spectra. Typically employed 1D non-LTE stellar simulations do not fully capture the complexity of stellar photospheres. Often only a fraction of the spectrum (the locations of a subset of cleanly identified features) is utilized. There may also be systematic abundance offsets in the derived abundance labels due to signal-to-noise dependencies of their derivation, or unmodelled instrumental imprints on the spectra, meaning that abundance estimates are subject to artefacts \citep[e.g.][]{SDSSdataproducts}. Data-driven approaches have provided higher precision abundances for stars across surveys \citep[e.g.][]{Ness2015, Ho2017, Casey2017, 
Wheeler2020}. However, these approaches still at their core rely on stellar models to provide stellar parameter and abundance labels for the training data.

In this paper, we demonstrate the feasibility of identifying chemically identical stars without explicit use of measured abundances. We apply a neural network with a supervised disentanglement loss term to a synthetic APOGEE-like dataset of spectra. The model learns a representation of spectra that traces abundances independently from the non-chemical factors of variation. That is, it controls for changes in the spectra caused by, for example, effective temperature, \teff, and surface gravity, \logg. This isolates the chemical variation expressed in the spectra.
Stars with identical chemical compositions but differing \teff\ and \logg\ are mapped to nearly identical representations. 

Unlike approaches based on explicit abundance estimates, this model naturally exploits the full available wavelength range including blended lines to estimate, effectively, chemical composition. Additionally, it does not depend on stellar models and so does not suffer from associated systematics. We find that the learned low-dimensional representation of synthetic spectra can be transformed linearly into abundances with high precision.

Our method relies on the assumption that there does not exist any correlation, nor statistical dependencies, between physical and chemical factors of variation. Although such an assumption has been used \citep{AbundanceReview,valenti2005},  stellar processes, such as atomic diffusion and dredge-up, contribute to modifying surface abundances away from their birth values \citep{AtomicDiffusionDotter}.  Because our model learns a representation of stellar spectra in which all variation dependent on non-chemical parameters is removed, assuming evolutionary changes in abundances correlate with the non-chemical factors we parametrize, the effect of these processes should also be removed from the representation, meaning that it will reflect birth, rather than present-day, abundances.

Studies of open cluster populations have demonstrated that stars can change in their element abundances by 0.1-0.3 dex across the main sequence to giant branch \citep{Souto_2019,Mota_2018}. This is also in line with theoretical expectations and a consequence of physical processes like atomic diffusion \citep{AtomicDiffusionDotter}. It is therefore a relevant and important distinction that we interrogate the spectra of a star for its birth abundance composition as opposed to its present day composition.

We have structured our paper such that our technical work on supervised disentanglement, of potential interest outside the astronomy community, is presented separately from our astrophysical application. After introducing the associated literature in Section \ref{sec:related}, in Section \ref{sec:problem} we discuss disentangled representation learning. In Section \ref{sec:application}, we adapt this method for chemical tagging and show our experimental results on an APOGEE-like dataset, demonstrating the recovery of chemically identical stars in the presence of noise. We also compare our approach to a baseline method \citep{Price-Jones2019}. We finish by discussing in Section \ref{sec:discussion} some important aspects of our method which are not explored using synthetic data, that comprise the next steps as well as our method's benefits. 

\section{Related Work}\label{sec:related}

\subsection{Disentangled representation learning}

There is a growing body of literature on using neural networks for learning to encode data into interpretable representations. Unsupervised disentanglement methods, such as beta-vae \citep{DBLP:conf/iclr/HigginsMPBGBML17} and infogan \citep{DBLP:conf/nips/ChenCDHSSA16}, attempt to find representations in which distinct informative factors of variation (such as lighting conditions and object orientation in the context of images) are encoded in separate dimensions \citep{Bengio}.  However recent results suggest that finding such disentangled representations in a fully unsupervised setting is fundamentally ill-posed without additional assumptions or priors being set \citep{DBLP:conf/iclr/LocatelloBLRGSB19}.

Supervised disentanglement methods \citep{Schmidhuber,GaninDomainAdaptation,DBLP:conf/nips/LampleZUBDR17} specify labels for factors of variations that should be excluded from the learnt representation. They aim to find a representation of inputs in which the specified factors of variation are removed from the representation but for which all other factors of variation are still present. A perfectly disentangled representation is statistically independent from the specified factors of variation. However, there will often be a trade-off between disentanglement and reconstruction \citep{tradeoff}.


Supervised disentangled learning has primarily been implemented through an adversarial training scheme, in which an autoencoder---a neural network with a lower-dimensional bottleneck that is trained at reconstructing inputs---learns to encode its input in such a way that a second network is unable to predict the to-be-disentangled labels from the encoded representation \citep{DBLP:conf/nips/LampleZUBDR17,DBLP:journals/corr/EdwardsS15,DBLP:conf/cvpr/HadadWS18}. It has also been proposed to obtain a disentangled representation by enforcing that an autoencoder learn a representation in which latent and labels are factorized. This has been done within the variational autoencoder framework  in \cite{DBLP:journals/corr/LouizosSLWZ15}, but also with adverserial autoencoders in \cite{polykovskiy2018entangled}. Another existing avenue for obtaining supervised disentanglement can be found through a cyclic training scheme, that encourages the latent to remain unchanged after re-encoding outputs, obtained after modifying the factors of variation. This approach has been demonstrated in the context of variational autoencoders in \cite{cyclicVAE1} and in \cite{cyclicvae2}.

Supervised disentanglement could be a very useful technique in the field of astronomy and we hope that this paper will be beneficial for showcasing its potential. For example, supervised disentanglement could be used in astronomical calibration to remove the effects of individual fibers or weather conditions on spectra. This could be done by learning a representation that is, for example, statistically independent from the fiber number for a  multi-object spectrograph. Such an approach would be complimentary to our paper, as our proposed method requires precisely calibrated spectra and requires additional augmentation to handle systematic artefacts.

\subsection{Data-driven chemical tagging} \label{sec:ddtagging}

Chemical tagging describes the reconstruction of individual cluster groups via abundance information \citep{TaggingFreeman, Ting2016, Casey2019}. The concept has extended to the identification of chemically anomalous stars of particular formation origins \citep{Hogg2016, Schiavon2017}, the association of and differentiation between stellar groups and populations using abundances \citep{Simpson2019, Hawkins2018, Martell2016} and grouping stars by chemical similarity \citep[e.g.][]{Price-Jones2019}.
Recent work indicates there is limited feasibility of chemically tagging stars back to their individual cluster origins using the $\approx$ 20 individual abundance measurements alone from resolution R=22,500 spectra \citep[e.g.][]{Ness2018}. Most approaches use the labels that describe the spectra and new approaches have improved the precision of these labels \citep{Ness2015, Leung2018, Ting2019}. Novel approaches to chemical tagging include those presented in \cite{Blanco-Cuaresma2018} and \cite{Jofre2017} which use techniques from the field of phylogeny and \cite{Price-Jones2019} who identify chemically identical stars without explicit use of abundances. The concurrent work presented in \cite{cyclestar} uses a machine learning algorithm loosely similar to ours for improving stellar abundance estimation.

The method proposed in \cite{Price-Jones2019}, which itself expands upon earlier work presented in \cite{10.1093/mnras/stx3198f}, bears some clear similarity to our work in that it uses a data-driven model applied directly to spectra to learn a representation in which undesirable parameters are removed.  They fit a polynomial model of the non-chemical parameters to every single wavelength bin. The residuals of this fit are then considered to only contain chemical information. They then  run a clustering algorithm on a compressed representation of the residuals obtained after principal component analysis to identify chemically-similar groups. However, as discussed in their paper, this method comes with some limitations. A polynomial fit may not be an optimally flexible functional form, particularly across a breadth of stellar evolutionary states (see for example \citealp{Ting2019}). As such, it is unlikely to perfectly remove physical parameters of variation from the residuals. Furthermore, by fitting non-chemical parameters in isolation, any joint dependencies between chemical and non-chemical factors of variation on spectral line strengths are ignored.  

\section{Methods}\label{sec:problem}
We present, here, an overview of our method. Subsection \ref{sec:setting}, introduces our underlying assumptions on the data-generating process, the problem we are trying to answer, and the broad-strokes of our method. In Subsection \ref{sec:autoencoding}, we dive deeper and present a neural network architecture for solving our introduced problem. In Subsection \ref{sec:disentanglement}, we present two different methods of enforcing a disentangled representation - a key component in our method.

\subsection{Problem statement}\label{sec:setting}

We consider a setup in which a dataset $X=\left \{ x_1,...x_n\right \}$ is observed. We assume the dataset to be generated deterministically from latent variables through a mapping unknown to us. Despite not knowing this mapping, we assume that a subset of the latent variables can be accurately estimated. As such we can subdivide latent variables into a vector of known variables $u$ and a vector of unknown variables $v$.  For our method to work, we further assume that  $u$ and $v$ are (marginally) statistically independent (i.e. $p(v|u)=p(v)$). This corresponds to the notion that $u$ and $v$ can not be used to predict each other.

In this paper, we present a general method for quantifying the similarity of observations $x$ as measured in terms of unknown variables $v$. In particular, our method provides a mean for identifying observations $x$ sharing identical or near identical vector $v$ without knowledge of the mapping from latent to observed variables.

Our method learns a mapping, parameterized by a neural network, from observations $x$ to a vector $z$ acting as a proxy for unknown variables $v$. More precisely, we learn a mapping such that observations sharing a common parameterization for $v$, in turn, share a near identical representation for $z$.

We achieve this through finding a representation $z$ which is statistically independent from the known and provided parameters, $u$, but when combined with these known parameters, capable of perfectly reconstructing observations $x$. This ensures that our latent variables contains all the information contained within the unknown variables $v$ but not any additional superfluous information.

How does this assumed setup relate back to astronomical chemical tagging? For chemical tagging, we have access to stellar spectra of stars, $x$, from which we seek to identify stars sharing an identical chemical composition, $v$. Although we are capable of estimating physical parameters, $u$, fairly accurately, shortcomings in spectral synthesis make it difficult to relate spectra back to their chemical composition.

\subsection{Approach} \label{sec:autoencoding}
\begin{figure*}
\includegraphics[width=\linewidth,trim={0 3cm 0 3cm},clip]{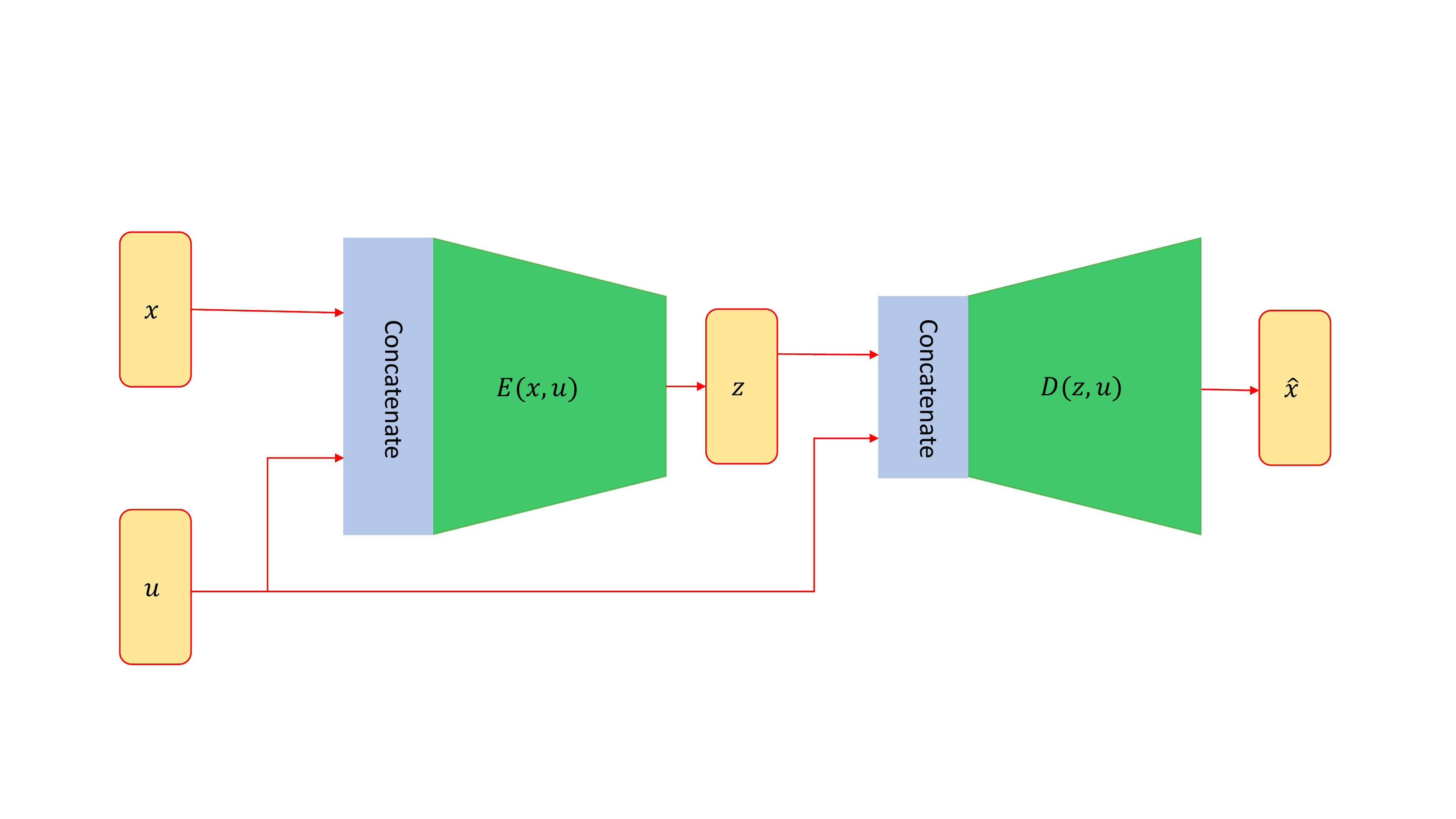}
\caption{Diagram of the conditional autoencoder architecture. We denote the reconstructed observation as $\hat{x}$. For chemical tagging, $x$ corresponds to stellar spectra and $u$ to phyiscal factors of variation.}
\label{fig:autoencoder}
\end{figure*}

We rely on a conditional autoencoder, a type of neural network, to learn the mapping to the lower dimensional representation $z$. Our autoencoder (represented in Figure \ref{fig:autoencoder}) is composed of two seperate neural networks. A conditional encoder taking as inputs observations, $x$, concatenated with known parameters, $u$, and returning a latent representation, $z$, (for the remainder of the paper we adopt machine learning terminology and refer to z as latents) and a conditional decoder taking $z$ and $u$ as input and trained to output reconstructed observations $x$.

This autoencoder is trained to minimize the following loss function:
\begin{equation}\label{eq:global_loss}
   L_{AE} = L_{rec}+\lambda L_{dis}
\end{equation}
where $L_{rec}$ is a reconstruction loss. In our experiments we used the mean squared loss 
\footnote{$\| x  \|_{2} := \sqrt{x_1^2+...+x_n^2}$ }

\begin{equation}
 L_{rec} = E_{(x,u) \sim p(x,u)}[\left \| D(E(x,u),u)-x \right \|^{2}_{2}]
\end{equation}

$L_{dis}$ is a disentanglement loss, acting to ensure that the latent, $z$, is maximally disentangled from the known and provided parameters, $u$. $\lambda$ is a term controlling the trade-off between reconstruction and disentanglement.The disentanglement loss is there to push the network towards learning a latent representation that is statistically independent from the observed parameters, $u$, and should be minimal when $z$ is independent from $u$. We present two formulations of $L_{dis}$ in Section \ref{sec:disentanglement}.

During training, the autoencoder is iteratively shown datapoints, grouped into batches -- subsets of the dataset. The autoencoder's loss, as described above, is evaluated on each batch and the derivatives of this loss with respect to the neural network parameters are used to update the parameters in the direction minimizing the loss function. After training, the neural network will have converged to parameterizing a mapping which (locally) minimizes the loss function. Although not a global minima, the learned mappimg, in part because of the stochastic nature of the training process, will typically be a good minimizer of the loss function.

Our neural network, in minimizing the loss function described by Eq. \ref{eq:global_loss}, simultaneously minimize reconstruction and disentanglement terms with a trade-off controlled by $\lambda$. Minimizing the disentanglement loss term corresponds to learning a latent representation statistically independent from factors of variation parameterized by $u$. This is achieved by removing all related information from the latent. The reconstruction term will be minimized when $z$ and $u$ are sufficient for reconstructing observations $x$. Combined, these two loss terms will be minimized when all the information required for modelling observations $x$ not included within $u$ is contained within the latent $z$. While it may not always be possible to minimize both loss terms together, we know that it is possible to do so for data generated as described in \ref{sec:setting}. Indeed, a global minimum of the loss function would be reached for a neural network which encoded observations $x$ into $v$ and decoded back to $x$.

In addition to isolating unknown factors of variation, $v$, we have found that, at least for the problems we have considered, supervised disentanglement maps observations with shared parameter values $v$, to nearly identical latents, $z$. We attribute this to our set of assumptions (see section \ref{sec:setting}).  This property makes some intuitive sense when we take a moment to consider how our autoencoder might map observations, $x$, generated from a common shared vector of unknown parameter values, $v$, but each with different values of the observed parameters, $u$. If the mapping does not project all of these observations to a common latent value, then the latent value, $z$, will be informative about the parameter value $u$ (as some $u$ are then more or less likely based on the observed $z$). Therefore, $z$ and $u$ will no longer be statistically independent.

In practice, our neural network will only approximately minimize our loss function and so will not perfectly map observations sharing common parameter values, $v$, to the same latent $z$. Observations sharing common parameter values will thus appear as over-densities in the latent space. These over-densities can then be identified, for example by running a clustering algorithm such as K-means \citep{Kmeans}, or by finding those observations particularly close according to some distance metric. Alternatively, we can instead identify such observations in the data-space if we use the decoder to convert all latents with a common set of parameters, $u_i$.

\subsection{Implementation of supervised disentanglement}\label{sec:disentanglement}
 
We present two alternative methods, FaderDis and Factordis, for learning a disentanglement loss $L_{dis}$ encouraging statistical independence. FaderDis is an adaptation of the Fader disentanglement architecture presented in \cite{DBLP:conf/nips/LampleZUBDR17} modified for our purposes. FactorDis is, to our knowledge, a novel architecture for supervised disentanglement.  We present here the architectures investigated. 

\subsubsection{Factor Disentanglement (FactorDis)}

The FactorDis method enforces independence by training a critic network to differentiate between samples from the joint distribution $p(x,u,z)$ and samples in which the statistical dependency between $z$ and $u$ has been forcibly removed. Analogously to generative adversarial networks \citep{Gan}, the conditional autoencoder is adversarially trained to generate samples that hinder the critic network's ability to do its job.

The joint distribution $p(x,u,z)$ can be expressed using Bayes rule as: 
\begin{equation}
\label{Joint}
    p(x,u,z)=p(x|u,z)p(u,z)
\end{equation}
This can be rewritten as 
\begin{equation}
\label{Factorized}
    q(x,u,z)=p(x| u,z)p(u)p(z)
\end{equation} if and only if $u$ is statistically independent from $z$. If the joint distribution $p(u,z)$ is not factorizable, the distributions $q(x,u,z)$ and $p(x,u,z)$ will be different. It follows from this, that $u$ and $z$ are statistically independent, if and only if samples $(x,u,z)$ drawn according to $p(x,u,z)$ are indistinguishable from those sampled from $q(x,u,z)$.

How can we generate samples from these two distributions? If we consider our autoencoder to be an idealistic autoencoder capable of perfectly reconstructing its inputs, then the encoder and decoder can be viewed as respectively approximately parameterizing $p(z|u,x)$ and $p(x|u,z)$, which are both deterministic functions. We can thus draw samples from $p(x,u,z)=p(z|u,x)p(u,x)$ by first randomly sampling from the dataset to obtain $(u,x)$, and then using the encoder to obtain the associated $z$. 

We can similarly draw samples from $q(x,u,z)=p(x|u,z)p(u)p(z)$ through reusing our samples drawn from $p(x,u,z)$. By scrambling $(u,z)$ pairs within a batch, we can effectively remove any joint information between $u$ and $z$ \citep{Mine} which results in samples drawn from the marginal distribution $(u,z) \sim p(u)p(z)$. We can then use the decoder which approximates $p(x|u,z)$ to obtain approximate samples $q(x,u,z)$ drawn from $p(x|u,z)p(u)p(z)$.

As stated above, enforcing statistical independence is the same as finding a latent representation, $z$, such that samples drawn according to these two procedures are indistinguishable. This bears strong similarity to the training objective of generative adversarial networks which attempt to train a generator such that generated samples are indistinguishable from samples drawn from a dataset. As such, we can take inspiration from existing generative adversarial network architectures to solve our disentanglement objective. 

In generative adversarial networks \citep{Gan}, a critic network is trained to distinguish between samples drawn from a dataset, and samples created by a generator network fed samples from a well-understood probability distribution. The generator and critic network are jointly optimized in a minimax game. That is, the critic attempts to maximally distinguish between the two data streams and the generator attempts to minimize the critic network's ability at doing so. The global optimum of this two player game occurs when both the generator and critic network can no longer improve - when the two data streams are identical.

For our disentanglement neural network architecture, we take heavy inspiration from Wasserstein generative adversarial network \citep{wgan}. We use an architecture parallel to that of generative adversarial networks. However, instead of differentiating between real and fake samples, we differentiate between samples from $p(x,u,z)$ and from $q(x,u,z)$ generated using the autoencoder (AE). This leads to optimizing the following minimax objective:
\begin{align}
\label{wgan}
    & \rm min_{AE}\rm max_{C \in \mathbb{D}}   \mathbb{E}_{(x,u,z)\sim p(x,u,z)}[C(x,u,z)]
    \notag \\
    -
    & \mathbb{E}_{(x,u,z)\sim q(x,u,z)}[C(x,u,z)]
\end{align}

where $\mathbb{D}$ is the space of 1-lipschitz continuous functions and $C(x,u,z)$ refers to a critic network that takes as inputs a vector in which observations $x$, latents $z$ and parameters $u$ are concatenated and attempts to differentiate between the different type of samples generated by our autoencoder.

The critic network attempts to maximize Eq. \ref{wgan}. In order to constrain the critic network to learn a lipschitz continuous function, we add a gradient penalty term, weighted by a constant $\lambda$, to the loss as was introduced in \cite{wgangp}. This leads to a critic loss function
\begin{align}
    & L_{critic}  =  
    \mathbb{E}_{(x,u,z)\sim q(x,u,z)}[C(x,u,z)]
    -
    \mathbb{E}_{(x,u,z)\sim p(x,u,z)}[C(x,u,z)]
    \notag \\
    & +
    \lambda
    \mathbb{E}_{(x,u,z)\sim r(x,u,z)}
    [(\left \| \nabla_{x,u,z} C(x,u,z) \right \|_{2}-1)^{2}]
\end{align}

where $r(x,u,z)$ is implicitly defined as sampling uniformly along straight lines between pairs of points sampled from the distributions $p(x,u,z)$ and $q(x,u,z)$. Further information about this sampling procedure can be found in \cite{wgangp}.

Our autoencoder, which plays the role of a generator network, is trained to minimize Eq. \ref{wgan} while simultaneously minimizing the reconstruction loss function:
\begin{align}
    & L_{AE}=L_{rec}
     +  \lambda_2\mathbb{E}_{(x,u,z)\sim p(x,u,z)}[C(x,u,z)]
    \notag \\
    & -
    \lambda_2 \mathbb{E}_{(x,u,z)\sim q(x,u,z)}[C(x,u,z)]
\end{align}.

This loss function combines the reconstruction loss that is traditionally used for optimizing autoencoders with a Wasserstein loss. In addition, unlike for generative adversarial networks, as both data streams are passed through the generator, they are both used for optimizing the generator. Training involves jointly minimizing the critic and autoencoder losses. The two different types of losses are weighted by a factor $\lambda_2$. Experimentally, we found that it was crucial to correctly set the factor $\lambda_2$, such that neither the reconstruction term nor the disentanglement term in the loss dominated over the other.

\subsubsection{Fader Disentanglement (FaderDis)} 

The FaderDis method of disentanglement follows the setup presented in \cite{DBLP:conf/nips/LampleZUBDR17} in which an autoencoder is adverserially trained to learn a latent representation from which an auxiliary network is incapable of predicting $u$. Since the method is designed to operate on discrete variables, we discretize our the parameter of space $u$ into $n$ equal sized bins. 

In this method, an auxiliary network, $A$, accepts latents, $z$, as inputs and outputs a vector of size equal to the number of discretized bin. It is trained using a cross-entropy loss to predict the probability of the corresponding $u$ vector falling in each of the n bins. The autoencoder $AE$ is then trained alongside this auxilary network. The autoencoder attempts to minimize the auxilary networks loss weighted by a factor $\lambda_1$ while also maximizing its own reconstruction loss. The autoencoder loss takes the form
\begin{equation}
 L_{AE}= L_{rec} - \lambda_1 (E_{(x,u) \sim p(x,u)} [-u_n log(A(E(x,u)))]  )
\end{equation}
where $u_n$ denotes the one-hot-encoding vector after the discretization procedure with the subscript $n$ referring to the bin in which the parameters $u$ fall.

The global optimum of this two player minimax game will occur when the autoencoder learns to reconstruct observations using a latent $z$ which does not contain any helpful information for the auxilary network. Since the auxilary network attempts to learn $p(u \mid z)$, this will occur when $p(u \mid z) = p(u)$ or equivalently when $u$ and $z$ are statistically independent.

\section{Application to Stellar Spectra}\label{sec:application}

We wish to learn a representation of stellar spectra that disentangles factors of variation of interest (chemical abundances) from the observed parameters, $u$.  We tested both methods with metallicity, [Fe/H], as a known and unknown parameter ($u = [\teff, \logg, [\mathrm{Fe}/\mathrm{H}]]$ and $u = [\teff, \logg$]).  After training, without any explicit knowledge of abundance labels, $v$, our neural network will find a mapping from observations, $x$ and parameters, $u$, to latents, $z$,  such that stars sharing a common abundance are mapped to nearly identical latents.

We demonstrate our method using a synthetic dataset described in Section \ref{sec:dataset_design}. The dataset is designed to mimic the spectral variability found within the APOGEE red-giant sample. This allows us to carry out a proof of concept for our method in an ideal and controlled environment, in which independence between chemical and physical parameters is guaranteed and for which we were certain to have accounted for all factors of variation. This is an important first step in demonstrating the viability and performance of our method.

We quantify the performance of our generative model with a chemical abundance twin recovery test, comparing to simpler models, that also remove factors of variation \teff, \logg\ and \feh. We do this for a number of signal to noise qualities. We note that the performance of our method, in practice, will be sensitive to any calibration or instrumental artifacts that are poorly modelled or not included as observed parameters. We also expect that the dimensionality of real data may be far lower than that of our synthetic library. This is because we do not restrict our realized abundances to the correlations observed in real stars. We therefore make only a comparative analysis of different modeling choices in recovery of abundance twins, rather than make a quantitative prediction of performance for real survey data.

\subsection{Simulated dataset}\label{sec:dataset_design}

For the creation of our spectra, we relied on the APOGEE package introduced in \cite{Bovy2016}, which wraps the Turbospectrum spectral synthesis code \citep{2012ascl.soft05004P} using ATLAS9 atmospheres \citep{2012AJ....144..120M}.  We created identically distributed training and test datasets, both containing 25,000 pairs of chemical abundance twins, sharing identical surface chemical abundances but differing stellar parameters. We generated our spectra assuming solar isotopes ratios.

When creating our spectra, the non-chemical parameters varied were the effective temperature $\rm T_{eff}$ and surface gravity, $\rm log\,g$. For each spectra, $\rm T_{eff}$ and $\rm log\,g$ were generated by sampling from uniform distributions as found in Table \ref{tab:parameters}. These parameter ranges were designed to replicate those of red-giant type stars which are the favoured type of stars for chemical tagging \citep{Hogg2016,10.1093/mnras/stx3198f}. Chemical abundances were generated by independently sampling log-metallicity ([Fe/H]), and log-element abundance enhancements ([X/Fe]), assuming Gaussian distributed values. Our Gaussian $1-\sigma$ standard deviations were chosen to roughly reflect those observed by the APOGEE survey. Exact values can be found in Table \ref{tab:enhancement}. These were determined from the 1-$\sigma$ element abundance dispersion in APOGEE's DR14 for each element, for red giant stars. By fitting separate one-dimensional Gaussians to the element abundance enhancements, we ignore any further correlations that may exist between elemental abundances. In doing this, we will overestimate spectral variability (i.e. dimensionality), which could lead to our chemical tagging predictions being overly optimistic, for the set of stars we consider in our tests. The absolute performance that we later report in recovery of abundance twins, is subject to the number of stars we are evaluating, as well as their density in chemical element abundance space and the dimensionality of the spectra itself. Therefore, it is only the comparative performance between the approaches we show that is relevant.

\begin{table}
\centering
\begin{tabular}{p{2.0cm} p{3.0cm} p{3.0cm}}
 \multicolumn{3}{c}{Parameter ranges} \\
 \hline
Parameter & Min  & Max\\
 \hline
$\rm T_{eff} [K]$   &   4000 &   5000\\
$\rm log\,g [dex]$   &   1.5 &  3.0\\
\end{tabular}
        \caption{Table containing the ranges used for uniformly sampling the non-chemical parameters of variation.}
        \label{tab:parameters}
\end{table}

\begin{table}
\centering
\begin{tabular}{p{1.5cm} p{3.0cm} p{3.0cm}}
 \hline
$[\rm X/Fe]$ & Mean $\rm [dex]$  & Standard Deviation $\rm [dex]$\\
 \hline
$[\rm Fe/H]$ &-0.13  &   0.24\\
$[\rm N/Fe]$  &     0.28 &  0.11\\
$[\rm O/Fe]$ &     0.03  &   0.08\\
$[\rm Na/Fe]$ & -0.05 &   0.38\\
$[\rm Mg/Fe]$ & 0.06 &   0.08\\
$[\rm Al/Fe]$ & 0.07 &   0.09\\
$[\rm Si/Fe]$ & 0.05 &   0.07\\
$[\rm S/Fe]$ & 0.05 &   0.07\\
$[\rm K/Fe]$ & 0.04 &   0.07\\
$[\rm Ca/Fe]$ & 0.02 &   0.04\\
$[\rm Ti/Fe]$ & -0.01 &   0.06\\
$[\rm V/Fe]$ & -0.01 &   0.11\\
$[\rm Mn/Fe]$ & -0.04 &   0.07\\
$[\rm Ni/Fe]$ & 0.02 &   0.04\\
$[\rm P/Fe]$ & -0.04 &   0.18\\
$[\rm Cr/Fe]$ & -0.01 &   0.06\\
$[\rm Co/Fe]$ & 0. &   0.15\\
$[\rm Rb/Fe]$ & -0.03 &   0.29
\end{tabular}
        \caption{Mean and standard deviation used when sampling the chemical factors of variation. Enhancements and metallicity are assumed to be Gaussianly distributed in dex.}
        \label{tab:enhancement}
\end{table}

\subsection{Implementation details}\label{subsec:methods}

For both FaderDis and FactorDis, we performed a manual hyperparameter search on the training dataset to select the best-performing model. Results for selected models are then shown on the test dataset.  We chose to set the latent dimensionality, $z$, to have dimension 20, slightly exceeding the number of varied abundances. A more comprehensive description of our neural network architectures can be found in Appendix \ref{appendix:training}. Our code is available on GitHub at \url{https://github.com/drd13/tagging-package}.

We also evaluated the performance of the model developed in (\citealp{10.1093/mnras/stx3198f, Price-Jones2019}; described in Section \ref{sec:ddtagging}), which we refer to as \emph{PolyDis} from now on. We use PolyDis with a forth-order polynomial, as was found to work best on training dataset.

To better simulate real data, we also evaluate our methods on a test dataset with added Gaussian noise. For a given signal-to-noise ratio (SNR), we add to every bin of the continuum-normalized spectrum zero-mean Gaussian noise with standard deviation $\rm \sigma = \frac{1}{SNR}$.
For FactorDis, results on the noisy test dataset are obtained by training models on data in which noise of order 1 percent (SNR=100) is added to every observation during training.
For FaderDis (and PolyDis), noise of order 1 percent was added to the training data and kept constant for every epoch of training. It was found that adding this type of noise to the training dataset led to worsened spectral reconstruction but improved isolation of chemical factors of variation. The worsened reconstruction can easily be attributed to overfitting to the noise. We do not have a clear explanation for why it led to improved isolation of chemical factors of variation.

\section{Results}
In this section we present a series of experiments comparing and contrasting the capacities of the different models. 

\subsection{Resolving power of latent representation to distinguish chemically identical stars} 

If non-chemical factors of variation have been perfectly removed from the latent $z$, stars sharing a common chemical abundance should in turn also share a common latent vector. As such, any difference in latent representation between chemically identical stars can be attributed to imperfections in the learned representation. Here, we use this to compare and contrast how well our considered methods isolate-out chemical factors of variation.

\begin{figure*}
\includegraphics[width=\linewidth]{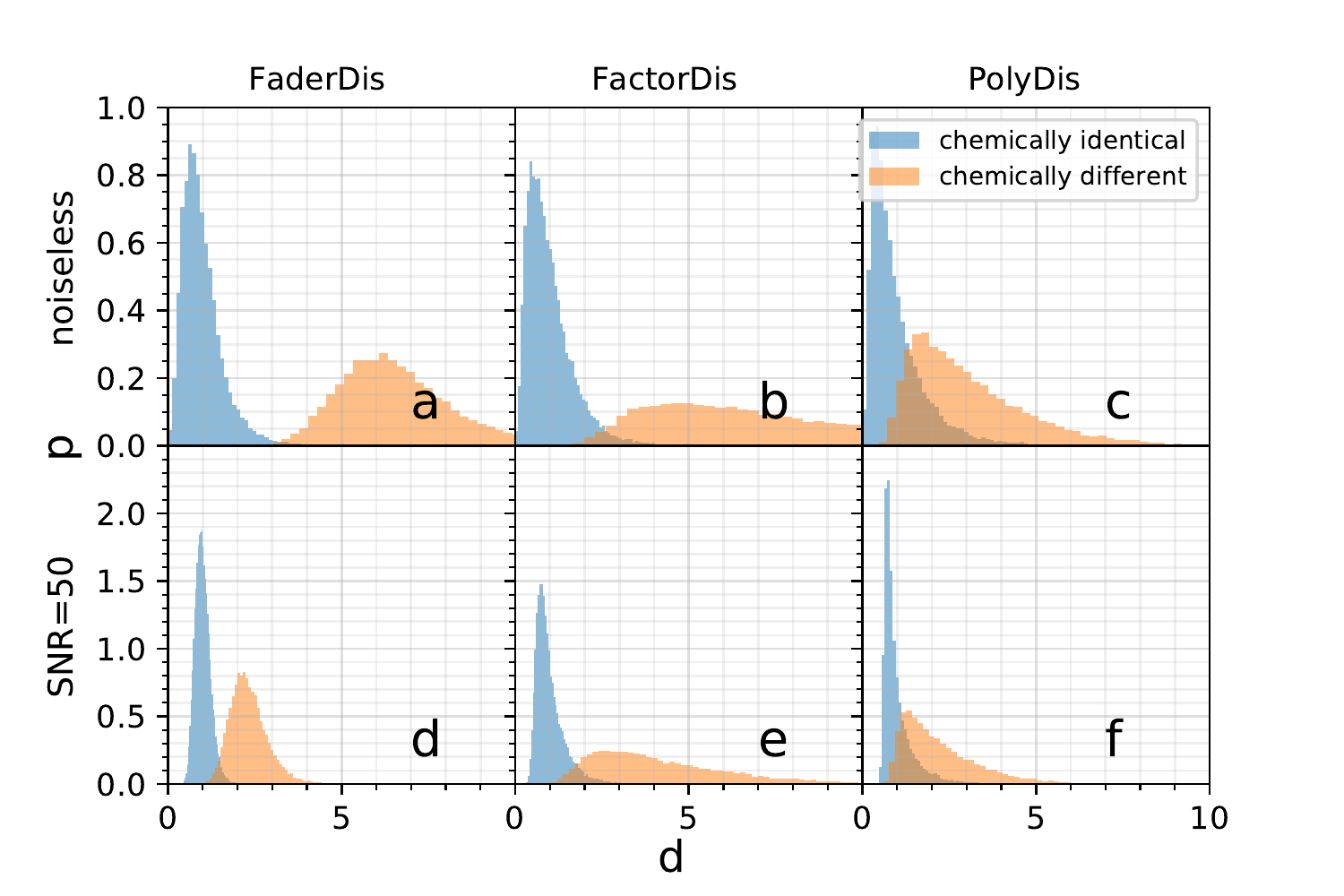}
\caption{Distribution of scaled euclidian distances, d,  for a sample of chemically identical pairs of stars (blue) and fully randomly sampled pairs of stars (orange). For each model, a scaling is applied to the latents such that the mean distance of chemically identical stars is 1. Each model includes \teff, \logg\ and [Fe/H], as the parameters to disentangle from the chemical factors of variation.  The top row is evaluated using the noiseless test dataset, the bottom with noise of order SNR=50 added. The first column is evaluated using the FaderDis method, the second using the FactorDis method and the final row using the PolyDis method (after PCA with 50 components).}
\label{fig:distributions}
\end{figure*}

Figure \ref{fig:distributions} shows histograms of euclidean distances, $\left \| z_{i}-z_{j} \right \|_{2}$, calculated on the latents, $z$, for both chemically identical pairs of stars (blue) and randomly sampled (non-chemically identical) pairs of stars (orange). We show this for our three different disentanglement methods at several SNR. Distances are evaluated on the 20 dimensional latent for both the FaderDis and FactorDis methods, and the 50 dimensional PCA components for the PolyDis approach. We found that 50 principle components explained $99.95 \%$ of the variance in the (noiseless) data. In the interest of making the comparison with PolyDis fair, we include [Fe/H] as a disentangled parameter during the training of the FaderDis and FactorDis methods.

Reassuringly, we find that for all considered methods, chemically identical stars share more similar latents than random pairs of stars. However not all methods are equally good at this task, with PolyDis underperforming compared to FaderDis and FactorDis. Indeed we find that, unlike the other considered methods, the PolyDis method has a non-negligible overlap between the distributions of chemically identical pairs of stars and of random pairs of stars. This means that for chemical tagging purposes, there will be a larger fraction of random stars in the dataset falsely appearing more chemically similar than genuinely chemically identical stars.

\subsection{Quantifying Chemical Tagging Performance}\label{subsec:tagging}

In this section, we evaluate the quality of our learnt representations directly on the task of chemical tagging. Since our dataset was designed such that every star has a unique chemical abundance twin, we can evaluate chemical tagging methods based on their capability at recovering these introduced chemical abundance twins. We once again use the euclidean distance in latent space  d = $\left \| z_{i}-z_{j} \right \|_{2}$ as our measure of chemical similarity between stars.

\begin{figure*}
\includegraphics[width=\linewidth]{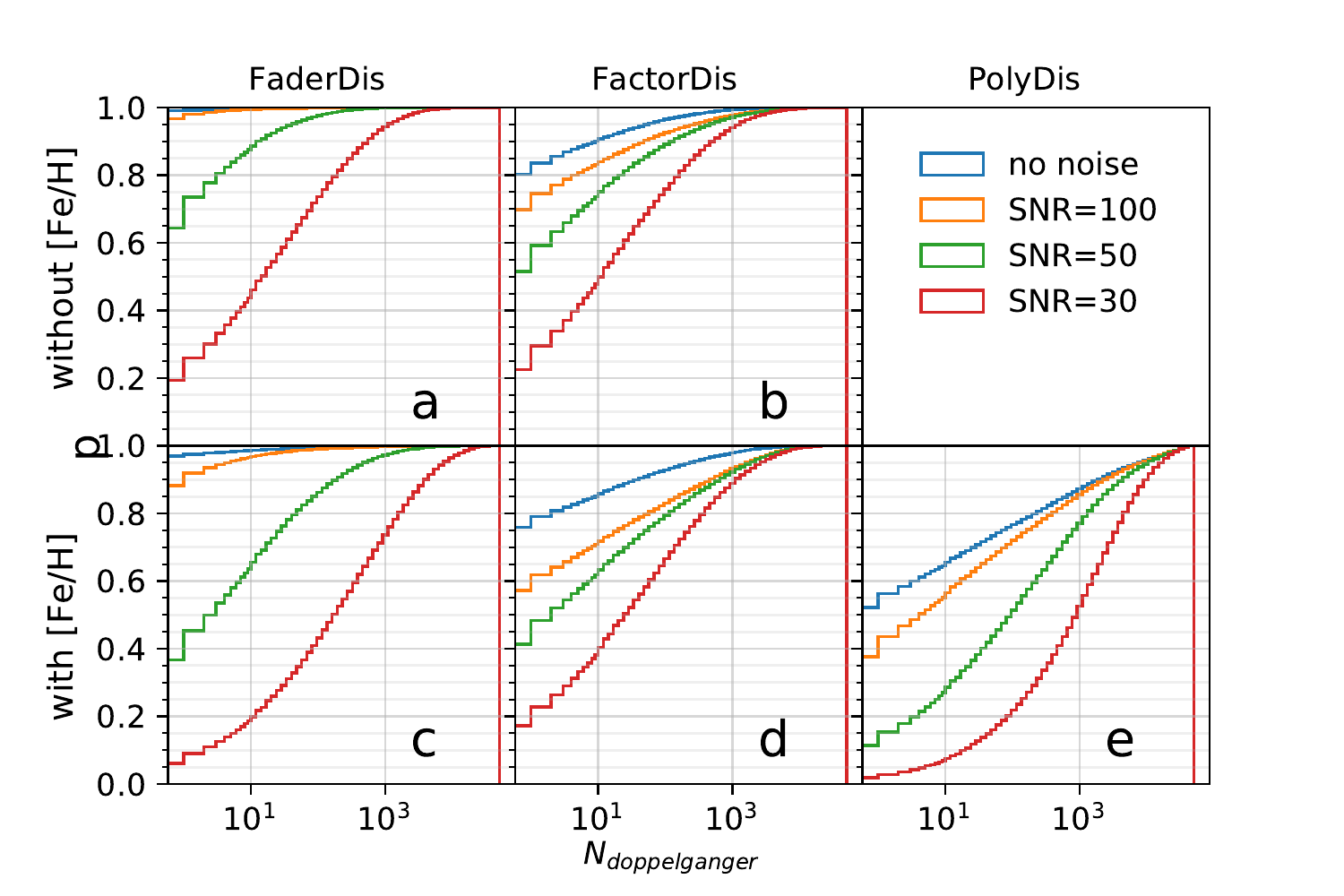}
\caption{Fraction of false chemical abundance twins for different models with differing signal to noise (SNR). In each panel, we plot the percentage of stars in the test dataset with fewer false twins than x, where x is the x-axis value, denoted as N$_{doppelganger}$. In the top row, we show results conditioned on \teff\ and  \logg. In the bottom row, we show results conditioned on \teff, \logg\ and [Fe/H]. We plot results obtained for FaderDis in the first column, with FactorDis in the second column and with PolyDis in the third column. It is worth reemphasizing that N$_{doppelganger}$ is highly dependent on the size of the dataset and as such this figure is only intended to be comparative and not as an absolute reference.}
\label{fig:cdf}
\end{figure*}

We show the results of our analysis in Figure \ref{fig:cdf}, where we have plotted the distribution of `false' chemical abundance twins recovered with each method - considered as stars  appearing more similar than the genuine chemical abundance twin. We term these our `doppelganger' stars. In our plot, the y-axis corresponds to the percentage of stars in the test dataset with fewer false twins then the corresponding value on the x-axis. For example when evaluating the FactorDis model that was trained to remove [Fe/H] on a dataset without noise (as shown in panel d), we found that around $85\%$ of stars in the dataset had fewer than 10 out of the 49998 other stars in the dataset being mistakenly measured as more chemically similar than their genuine chemical abundance twin. Similarly, the y-intercept represents the percentage of stars for which none of the 49998 other stars in the dataset are more similar than the genuine chemical twin.

The Figure suggests that precision disentanglement and removal of non-chemical factors of variation from stellar spectra is valuable for chemical tagging pursuits. The FaderDis method identifies significantly more pairs of chemically identical stars than the baseline PolyDis method. For example the FaderDis method, applied on a noiseless dataset with [Fe/H] removed from the representation (panel c), identifies around $97\%$ of pairs of chemical abundance twins compared to only $50\%$ for the baseline PolyDis method (panel e). For spectra with $SNR=100$, this number goes down to about $88\%$. As the neural network performance is sensitive to hyperparameters, architecture and loss function, any improvement in these areas could further improve results. For example, the FaderDis method was found to perform significantly worse when noise was not added as described in the implementation details.

Note that as we randomly generate our stars from a high dimensional distribution, there is a possibility for random pairs of stars to be chemically similar by chance. However we expect a chance of no more than 10$^{-12}$ of doppelganger pairs given the high dimensionality of the artificially generated dataset. Even if these exist however, our Figure is comparative only, to demonstrate how the three different methods work to recover the designated chemical abundance twin stars. As we are generating data from a fixed range of 20 independent abundance labels, recovery of chemical abundance twin stars will become harder under various conditions. This includes as the size of our test set grows within its current abundance ranges, and if correlations between the abundances were included in their prescription. We highlight, however, that this is a comparative test, to demonstrate the relative performance of the three methods, and as a function of signal to noise. The absolute performance would vary in the physical abundance distribution plane of real data.

\subsection{Interpretability of latent representation}

In this section we investigate whether the latent representations that organically emerges from our neural networks are interpretable. As our encoder and decoder are non-linear functions we might pessimistically expect our latent representations to be non-interpretable. We show that this is not the case and instead that, at least on our synthetic dataset, the learned representations align well with the measured abundances.

We approach this question through learning a linear transformation converting from latents to abundances. We represent our dataset of abundances and latents as matrices $V$ and $Z$ of shape $n_{species} \times n_{data}$ and  $n_{z} \times n_{data}$ where $n_{species}$ is the number of chemical species in the spectra, $n_{z}$ is the dimensionality of the latent space and $n_{data}$ is the number of observations in the dataset. We seek a transformation matrix $A$ converting latents into abundances as faithfully as possible. We can find such a matrix by solving  $\argmin_A || AZ - V||^2$ which has known solution $A = VZ^{+}$ \cite{matrixCookbook} with $Z^{+}$  the Moore-Penrose inverse of $Z$. We solve this matrix using all stars stars in the noiseless training data.

\begin{figure*}
\includegraphics[width=\linewidth,height=0.74\textheight]{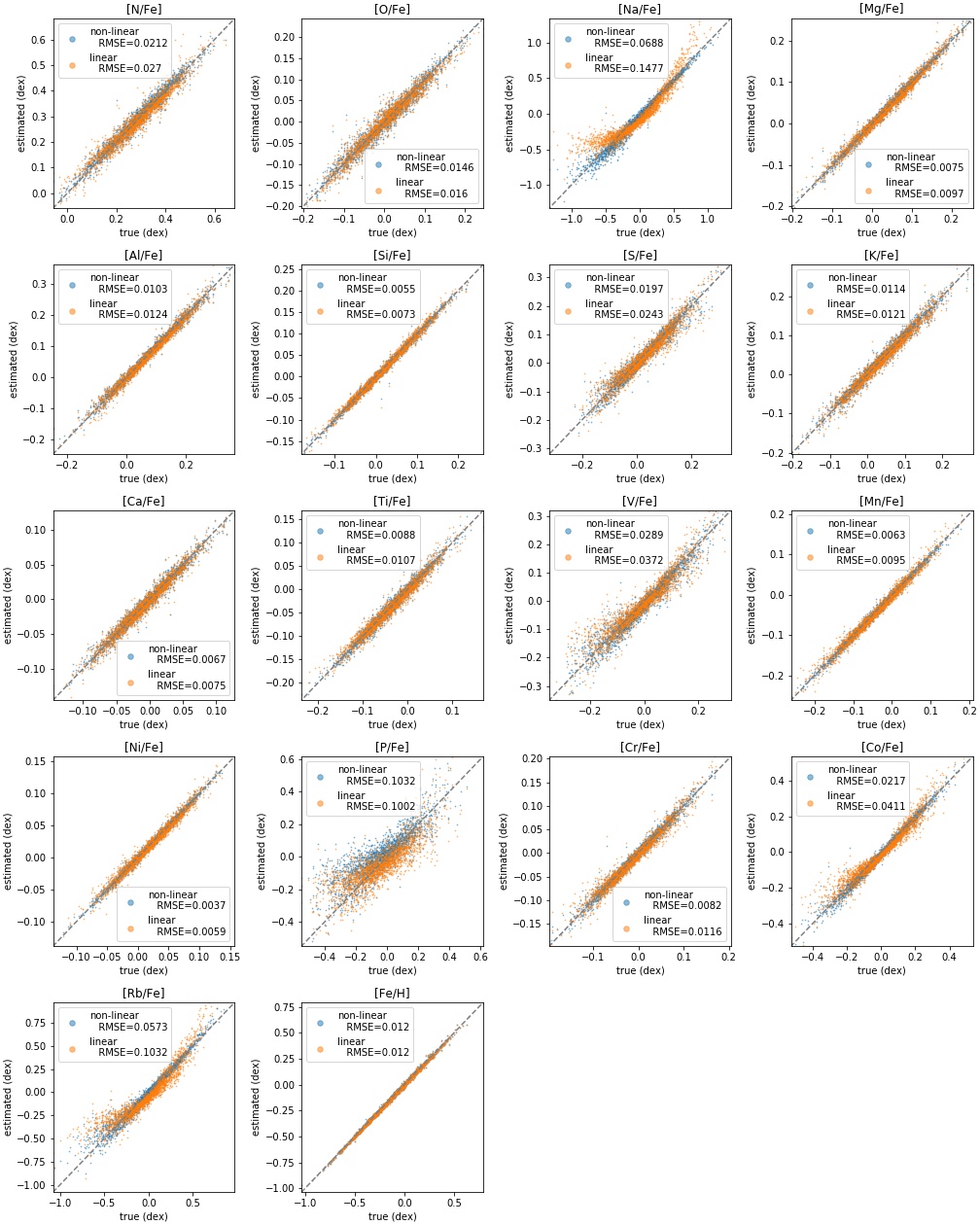}
\caption{Scatter plot showing estimated against true chemical enhancements and metallicities for synthetic stars in our test dataset. In the legend, linear refers to abundances estimated by multiplying the latent with matrix A, and non-linear to abundances estimated from the latent using a neural network. This figure was obtained using the latent from a FaderDis model trained at disentangling [$\rm T_{eff}$, $\rm log\,g$]. For each chemical element, we have also estimated the root-mean-square error (RMSE), the standard deviation of the residuals between predicted and true enhancements/metallicity.}
\label{fig:interpretability}
\end{figure*}

In Figure \ref{fig:interpretability} we have plotted chemical compositions as estimated from the linearly transformed latents against true chemical compositions. These are shown for 2000 stars in the noiseless test dataset. We see a remarkable agreement between the estimated and true abundances. For almost all species, the linear transformation is nearly as good at estimating chemical compositions as a neural network trained on the latents (denoted ``non-linear") on the same stars. Although Na is not as well fit as other species, it is known to be particularly difficult to estimate \citep{Ness2019,J_nsson_2018}. This shows that our method has naturally learned to decompose spectra into a representation nearly equivalent to chemical abundances. Although these results were obtained on a synthetic dataset they are particularly encouraging. Measuring abundance variation quantitatively, without reliance on synthetic spectra would allow for fully circumventing the uncertainties propagated from inaccuracies in spectral modelling.

\subsection{Spectral Reconstruction}\label{subsec:reconstruct}

Our neural network encoder allows for converting spectra into a representation in which predefined non-chemical factors of variation are removed. By subsequently applying the decoder to this representation, we can generate modified spectra recast to new non-chemical parameters.

\begin{figure*}
\includegraphics[width=\linewidth]{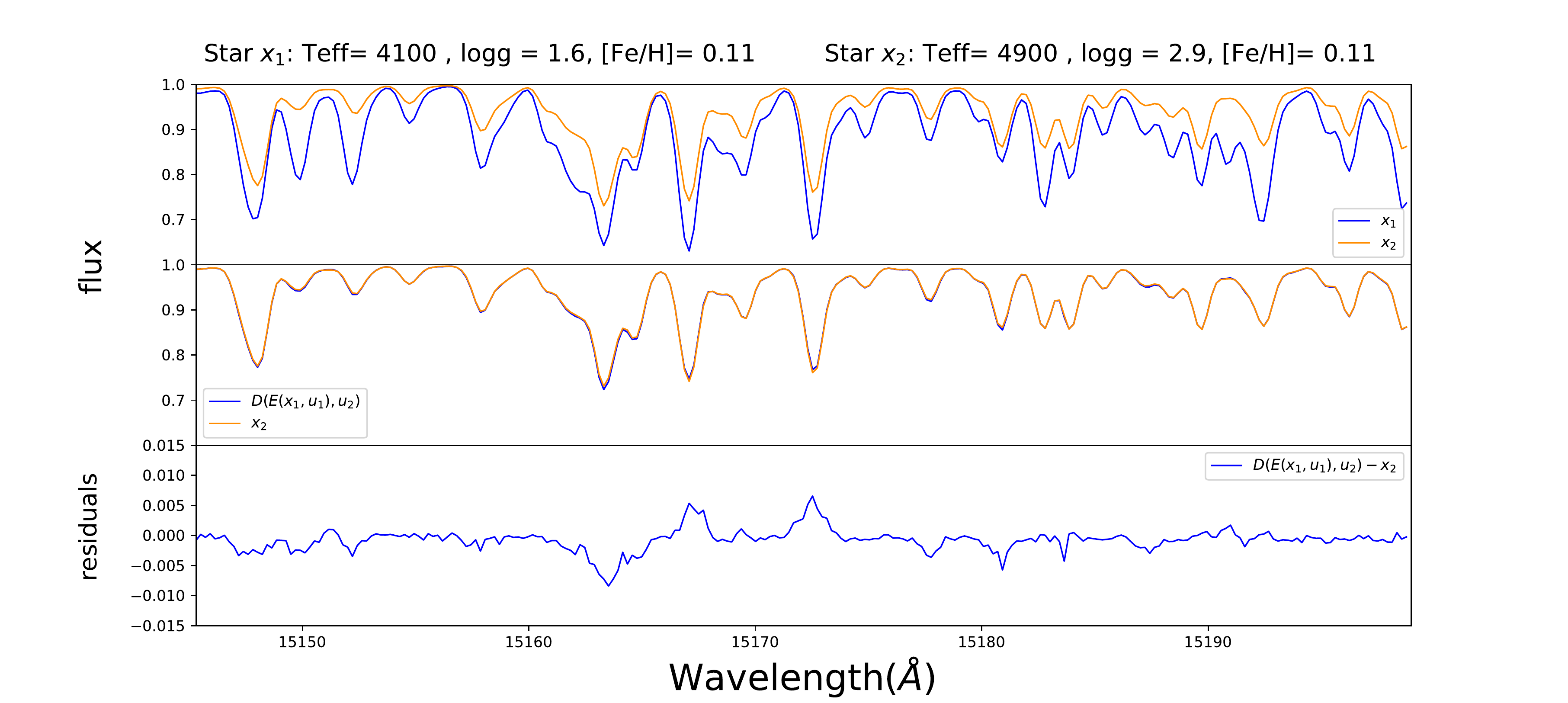}
\includegraphics[width=\linewidth]{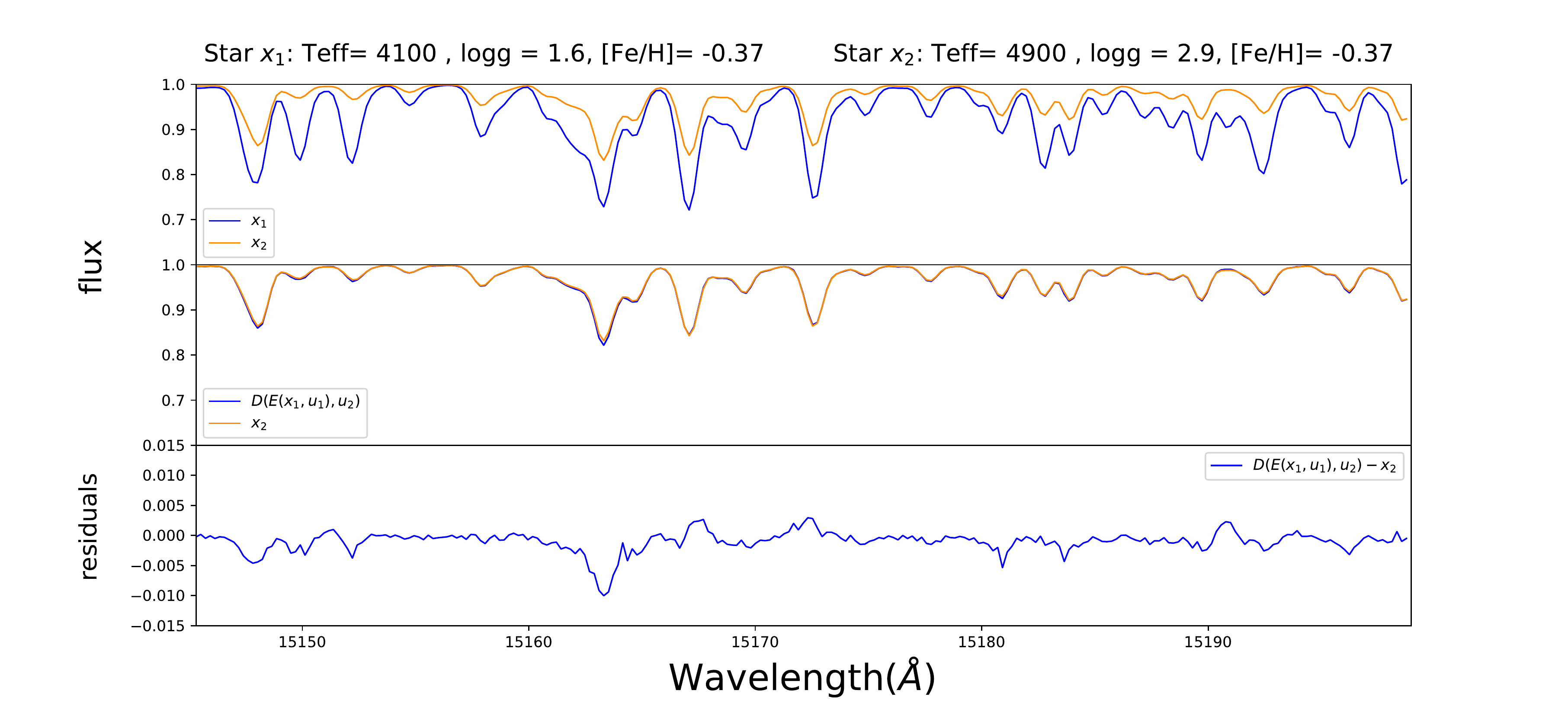}
\caption{For each subfigure, in the top panel, we show the spectra of two stellar chemical abundance twins (with differing \teff\ and \logg), $x_1$ and $x_2$. In the middle panel, the spectra of the second chemical abundance twin, $x_2$, is shown with a spectra reconstructed by the decoder ($D(E(x_{1},u_{1}),u_{2})$) using the other star's latent $z_1$ but the same physical parameters $u_2$. In the bottom panel, the corresponding residuals. The stellar parameters are shown above each subfigure (For conciseness The [X/Fe] vector is not shown). We can see that the spectra of chemical abundance twins are nearly indistinguishable after transforming them to a common physical parameter (\teff\ and \logg) parameterization.}
\label{fig:results}
\end{figure*}

In Figure \ref{fig:results}, we leverage this to visually demonstrate, for the FactorDis approach, how well our learned representation isolates the chemical information in the spectra of a pair of metal-rich stars (top plot, stars $x_1$ and $x_2$) and a pair of more metal-poor star (bottom plot, stars $x_3$ and $x_4$). These test spectra have been generated as described in Section \ref{sec:dataset_design}, with each pair sharing identical chemical compositions but differing physical parameters. 

For each sub-figure, in the top panel we plot the original pair of stellar spectra. In the middle panel, we plot how these same chemical abundance twins appear after $x_1$ is transformed to the physical parameter of $x_2$, and in the bottom panel we plot the residuals between the twins after the transformation. From these Figures, we see that although the initial spectra are very different, the transformed spectra are near identical. This is because the encoder isolated the chemical information and the decoder generated the recast spectra (for star $x_1$), at the new provided physical parameters (of the star $x_2$).

\begin{figure*}
\includegraphics[width=\linewidth]{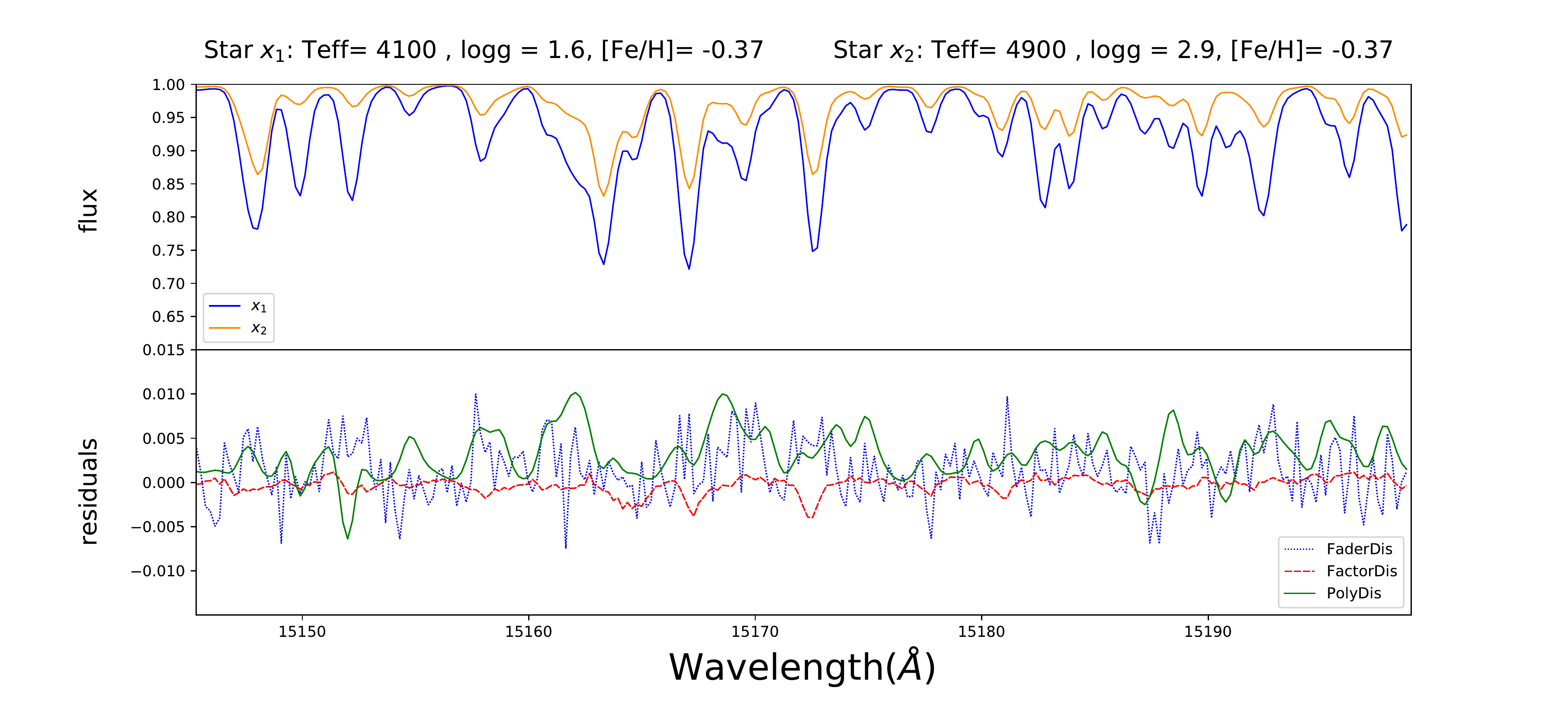}
\caption{This Figure compares the reconstruction capacities of the three disentanglement methods for the metal-rich star shown in Figure 2. In the top panel, the spectra of two chemical abundance twins, $x_1$ and $x_2$, for the first 256 wavelength bins. In the bottom panel, the residuals between the second twin, $x_2$, alongside the spectra of the first twin $x_{1}$, recast by the decoder ($D(E(x_{1},u_{1}),u_{2})$) to the physical parameters $u_2$ for the three disentanglement methods considered. The mean residuals and associated standard deviation (per pixel across the full spectral range) are $\rm R = 0.0029$ and $\sigma_R = 0.0021$ for FaderDis, $\rm R = 0.0011$ and $\sigma_R = 0.0009$ for factorDis and $\rm R = 0.0034$ and $\sigma_R =  0.0023$ for polyDis.}
\label{fig:residuals}
\end{figure*}
 

%

In Figure \ref{fig:residuals}, we show the residuals between a star and its transformed twin for FactorDis, FaderDis and PolyDis. For this comparison, we include the three factors of non-chemical variation, \teff, \logg\ and \feh, in the disentanglement network training. Alongside the Figure, we also report the mean absolute residual across the full spectral region considered as well as the standard deviation (per pixel) of the residuals, $\sigma_R$ for each approach. For the PolyDis method, a star is recast to its chemical abundance twin star's stellar parameters, by replacing its residuals from the polynomial fit with those of its twin star (the fit is meant to isolate the chemical information into the residuals).

In Table \ref{tab:quantitative}, we report the average mean absolute residual $\rm \langle R\rangle$ and average mean squared error $\rm \langle MSE\rangle$, obtained by averaging over random pairs of chemically identical stars in the dataset, transformed to each others physical parameters. The $\rm \langle MSE\rangle$ metric more severely penalises large deviations in the reconstructed compared to original spectra. Several interesting trends appear in the data.

We observe a difference in performance between methods, depending on whether the residuals or the squared residuals are used for evaluation. Most notably, the FactorDis method outperforms the PolyDis in terms of squared residuals but not raw residuals. As squared values are more sensible to outliers, this seems suggestive that the PolyDis method has comparative better overall reconstruction but struggles with representing some portions of the dataset. 

\begin{table}\small
\centering
\begin{tabular}{p{2.5cm}p{1.5cm}p{2.0cm}}
 \hline
Method & $\rm \langle R\rangle$ & $\rm \langle MSE\rangle$ \\
 \hline
FactorDis   &   0.0021 & 1.26 $\times$ 10$^{-5}$\\
FaderDis   &   0.0030 & 1.68 $\times$ 10$^{-5}$\\
PolyDis   &   0.0018 & 1.50 $\times$ 10$^{-5}$ \\
\end{tabular}
        \caption{Average MSE between two chemically identical stars transformed to each others physical parameters for the different methods. The quoted number assumes a dataset of stars distributed following the procedure as described in Section \ref{sec:dataset_design}.}
        \label{tab:quantitative}
\end{table}

The relative mean values reported in Table 3 are also largely indicative of the distribution in parameters of our library of test spectra that we have generated. Simpler modeling approaches likely perform very well when both the training and test data is smaller in overall variability and for pairs of stars that have nearer \teff\ and \logg\ parameters. In this case, the spectral variability due to the parameter and abundance labels is nearer to a linear or low order polynomial form \citep[e.g.][]{Ness2015, Casey2016}. In the regime that the stars considered cover a wide range in \teff\ or \logg\ and pairs of stars have much larger differences in these parameters, the move to more complex models, (or as an alternative, local linear models that build non-parametric models using nearest neighbours \citep[e.g.][]{Wheeler2020b}), may have higher return. These differences can also be understood in terms of the differences between methods at reconstructing spectra. In the PolyDis method, spectra are recast to new physical parameters, by adding residuals to the polynomial fit. This transformation is not parametric in the traditional sense, and so, if two stars being compared are similar to begin with, will give a very small reconstruction loss. For the case of pairs of identical spectra, this would give a perfect reconstruction, even if the residuals do not capture the chemical information. On the other hand, the FactorDis and FaderDis method involves decoding from a lower dimensional representation, and so, even for identical stars, have non-zero residuals.  The difference thus boils down to FactorDis and FaderDis being, by design, built for capturing chemical information, but not always (for our exercise) at reconstructing the stellar spectra while the PolyDis method can "cheat" at reconstructing stars. The FaderDis method does not perform particularly well at this task. We believe this to be linked to its training procedure which involves reconstructing noisy rather than clean data.
\begin{table}[h!]
\small
\centering
\begin{tabular}{p{4.0cm}p{2.0cm}p{2.0cm}}
 \hline
Method & $\rm \langle R\rangle$ & $\rm \langle MSE\rangle$ \\
 \hline
FactorDis   &  0.0027 & 2.13 $\times$ $10^{-5}$\\
FaderDis   &   0.0033 & 2.14 $\times$ $10^{-5}$\\
PolyDis   &   0.0029 & 3.33 $\times$ $10^{-5}$ \\
\end{tabular}
        \caption{Average reconstruction between two chemically identical stars transformed to each others physical parameters for the different methods on a \textbf{restricted} dataset composed of stellar chemical abundance twin pairs with at least 500K of temperature difference.}
        \label{tab:quantitative_restricted}.
\end{table}
To demonstrate that our method performs better on chemically disimilar stars, we have recalculated our metrics on a dataset restricted to stars with high temperature differences (see Table \ref{tab:quantitative_restricted}). On this partial dataset, the FactorDis method performs better across the board, and the PolyDis method performs worse than both FaderDis and FactorDis, in terms of $\rm \langle MSE\rangle$.

\section{Discussion}\label{sec:discussion} 

We have developed a neural network architecture to remove those factors of variation in stellar spectra that we want to disregard, from those others we care about. Here we want to isolate chemical abundances alone. Typically chemical abundances are measured from stellar spectra, which relies on imperfect stellar models and does not fully utilize the the full amount of information across the entire spectral region. We seek to develop approaches that circumvent limitations in our current knowledge of stellar physics or incomplete models, and leverage large surveys, by working as closely to the observed data space as possible. 

We compared two deep learning approaches and a simpler polynomial model approximation for the task of removing \teff, \logg, and also \feh\ from model mock stellar spectra -- leaving behind the intrinsic variation caused by chemical abundances. All three approaches perform well at generating a disentangled representation of spectra. A reader might note that the mean residual of the test data compared to the model, for all three methods we investigate, is lower than a typical Poisson noise level in an observed spectra in large surveys (often SNR $\approx >$ 50-100 per wavelength). The value of this level of precision, and importance of maximising the precision, comes when working with large stellar ensembles. Galactic archaeology typically demands large numbers of stars where the sampling precision of the population increases as (N)$^{1/2}$. The sampling error of the population itself at each wavelength, becomes smaller than the reconstruction precision, for hundred thousand star surveys.

We demonstrated the benefits of using a disentangled neural network,  using a chemical abundance twin star recovery, from our 50,000 star test set. This is related to the pursuit of chemical tagging, which is an extremely challenging aspiration with galactic archaeology---to find stars born together using their identical abundances. Even if chemical tagging is prohibited by field contamination, there is tremendous promise in modeling the distribution of the most chemically similar stars as a function of orbital properties or abundance density in reconstructing galactic history \citep[e.g.][]{Kamdar, Coronado2020, Price-Jones2020}. When applying our FaderDis approach to a synthetic dataset of APOGEE-like stars, we were able to identify the most chemically identical stars from the ensemble, even with moderate signal to noise. At an SNR of 100 we were able to identify around $87\%$ of pairs of stars. This compared to around $60\%$ of stars for our second neural network approach, FactorDis, and around $40\%$ for our implementation of a polynomial representation to subtract stellar parameters,  PolyDis, at this SNR.

Our results obtained using synthetic, model spectra are particularly promising. However, these  may not translate directly to real survey data, for a number of reasons. As our dataset was handcrafted, we were able to ensure our dataset perfectly matched the stringent requirements of our method. That is to say, we ensured perfect knowledge of all non-chemical factors of variation, and expressed these using a determinisitic parametrization, that is statistically independent from chemical factors of variation. We examine here if these assumptions are accurate for actual stellar surveys, and if not, how we might be able to modify our method to accommodate these discrepancies.

\subsection{Assumptions about stellar spectra}

Our method involves removing all non-chemical factors of variation. If our neural network is conditioned on an incomplete set of non-chemical factors of variation $u$, our latent will be contaminated by these when isolating chemical factors of variation. For actual observations, these nuisance parameters may arise from imperfect calibration such as from telluric lines or persistance in the detector or any other of a number of systematics. In principle, we may be able to somewhat counteract this phenomenon by restricting the dimensionality of our latent. This would force the latent to only encode the most important factors of variation. However, as some abundances only have a minuscule impact on the overall recorded spectral flux, we require very good knowledge of our factors of variation. An alternative approach for accounting for these systematics would be to add a disentanglement term targeting them. However,  this would require additional infrastructure not built into this first demonstration of the approach.

In our proof-of-concept experiments, we first modelled stars using only the effective temperature $\rm T_{eff}$ and surface gravity $\rm log\,g$ as non-chemical factors of variation. These two parameters should explain most of the non-chemical variance in the spectral data. Indeed, many data-driven models have been capable of accurately reconstructing spectra using these parameters, plus overall metallicity [Fe/H],  \cite[i.e.][]{Ness2015,Ting2019,Leung2018}, as these are responsible for the majority of spectral variability. However, other parameters, that are independent of the chemical composition, may also impact the observed spectra, and so may need to be included in our conditioning parameters $u$. Stellar mass, or age, for example, while correlated with effective temperature and surface gravity \citep{10.1093/mnras/stx3198f}, contains additional independent predictive power for generating the spectra \citep{MassCannon}. If this is indeed the case, it may be beneficial to include an independent estimate of stellar mass. This could be achieved by using a training dataset of stars from astroseismology surveys, with mass estimates. Stellar rotation may also affect stellar spectra. These variations may, at least in part, be captured by the micro and macro turbulence parameters.

Beyond assuming knowledge of non-chemical factors of variation, we have also so far assumed the ability to perfectly estimate these, if known. In realistic scenarios, this may not be easy. However, similarly to other data-driven methods such as \cite{MassCannon}, our method requires precise but not necessarily accurate parameter values. For example our neural network method should still be effective if a change of variable is applied to any of the conditioning variables. Furthermore, we do not account for the correlations between elements when we generate our test data, which will reduce the dimensionality of the spectra and effective sparsity of the data space.

Finally, even if we are unable to fully remove non-chemical factors of variation from spectra, our neural network architecture may still be useful for traditional chemical abundance estimation. Indeed, we may be able to reduce systematic uncertainties in traditional abundance estimation methods by recasting stars to a common temperature and surface gravity (as shown in section \ref{subsec:reconstruct}) before comparing to synthetic stellar spectra. Similarly to differential analysis, this would serve to restrict the number of factors of variation changing in stellar spectra.

\subsection{Assumptions relating to statistical independence}

Our approach assumes that abundances are statistically independent from other factors of variation. In our experiments, the synthetic spectral dataset was generated so as to satisfy this assumption. This assumption is not entirely unreasonable. There is evidence that most stellar abundances should, at least to first order, be independent from temperatures and surface gravity \citep{AbundanceReview}. Trends between abundances and physical parameters have, in the past, been attributed to systematic uncertainties and sometimes even been corrected for \citep{valenti2005,adibekyan}. However, overall metallicity does, at some level, affect stellar evolution \citep[e.g. see][]{Gaiadr2}. Ultimately this assumption breaks down to some degree and there is some level of statistical dependency between metallicity and physical parameters in observed spectra. Including overall metallicity in the disentangled parameters, as done in our experiments, mitigates this issue. Spectral synthesis approaches, including at low resolution, typically derive a basic set of \teff, \logg\ and [M/H] (or [Fe/H]) parameters (with errors), and so all parameters, including metallicity, are readily available to use in the disentanglement architecture we have built. We might also identify chemically identical stars by finding those stars sharing both a common latent representation and a common metallicity. Ultimately, accounting for any dependencies between abundances and parameters will better disentangle abundance variations from stellar parameter variations in real spectra. Large data sets may be leveraged to learn these dependencies. Indeed, stellar processes like dredge-up and diffusion \citep{Masseron2015, Martig2016} modify surface abundances away from their birth values across evolutionary state. Removing any trends caused by these processes would result in a chemical representation closer to birth abundances, which is ultimately preferable for using abundances for chemical tagging pursuits.

\subsection{Beyond synthetic spectra}

There are a few challenges associated with applying our method to real observations.  Spectral bins in real observations are sometimes flagged as untrustworthy, for example due to cosmic-rays or persistence in the detector (see \cite{J_nsson_2020}). These are flagged for each APOGEE spectra and individual pixels are correspondingly masked.  Since our neural network methodology requires all spectral bins as inputs, such untrustworthy or missing data need to be imputed somehow, so as to not impact the downstream learned representation. Another more practical challenge with applying the method to real observations is that the method requires hyperparameter tuning and training the algorithm takes a day to run on specialized hardware (GPUs). This makes iterative deployment slow.

\section{Conclusion}

Organising stars by their chemical similarity and investigating the distribution of their other properties (e.g. orbits, density) is a promising avenue for unravelling galactic evolution \citep[i.e.][]{Coronado2020, Kamdar, Ting2016}. Ranking stars by chemical similarity requires  precise chemical information for large numbers of stars. Chemical similarity is typically determined using measured element abundances. However, these measurements are subject to inaccuracies and systematics that are inherited from incomplete and approximate stellar models. As an alternative to deriving abundances from spectra, using the variability of the spectra itself, becomes possible and advantageous in the regime of large stellar surveys.  Data-driven deep learning methods - applied directly to the spectra itself - find natural applications here. 

In this paper, we introduce a new deep-learning method for extracting chemical information from spectra. This relies on isolating chemical factors of variation from non-chemical factors of variation, through training a neural network using a disentanglement loss. This method removes the need for accurate and precise modelling of the chemical abundance factors of variation in stellar spectra. Instead, it relies only on the parameterization of the other primary sources of variability, namely stellar parameters, including \teff\ and \logg. This requires conditioning a neural network on these factors, in our case, \teff\ and \logg, (and also \feh\ for modeling only the variation from abundance enhancements, [X/Fe]). We have shown, using a synthetic set of spectra that we have generated, that our method can be used to accurately identify and distinguish chemically identical pairs of stars from a field distribution, which is an aim of chemical tagging. We were able to identify more than $85\%$ of pairs of chemical abundance twins from a dataset of 50000 spectra, generated at the resolution of APOGEE and SNR of 100, with 20 independently drawn chemical abundances ([X/Fe]). To do this in practice (on real data), will require being able to estimate all non-chemical factors of variation in the spectra very well. In our analysis, as we wanted to demonstrate our method on a toy dataset with the fewest assumptions possible, we have treated the metallicity as statistically independent from physical parameters. For real observations however, it may be beneficial to account for their statistical dependency. As at fixed metallicity, chemical and physical parameters are believed to be independent or close to, we suggest to learn a representation in which metallicity is disentangled.

We note that our approach may also find utility in regimes where analysis is hindered by a large number of molecular bands in the spectra with uncertain atomic transition data that renders stellar models inaccurate. Our tests here are confined to a very narrow range of $\teff$ that is not dominated by molecular features but this may be a promising avenue for large survey data like Sloan V that will observe bright cool giants with molecular features \cite{SloanV}.

In this paper we have demonstrated,  through experiments on a synthetic dataset, the efficacy of our newly proposed method for extracting chemical information from stellar spectra. These experiments act as a proof-of-concept in a controlled environment, where the data generating process is perfectly understood. This sets out the groundwork for applying such representation learning methods to real observations. The natural next steps of this line of work will be translating the success found on synthetic spectra  to real APOGEE observations. Because of the imprint of systematics on real stellar spectra and other possible departures from our assumptions, this will likely require further modifications and/or fine-tuning of the approach. Such investigation are reserved for a future paper. In a forthcoming paper \citep{MetricDeMijolla}, we demonstrate on real APOGEE stellar spectra a method similar but complimentary to this one for extracting chemical information in a manner that is robust to instrumental systematics.

Beyond our astronomical contributions, we hope that our proposed methodology will find uses in other fields. Our application of supervised disentanglement for identifying observations sharing a common parameterization is a novel method that could be adapted to other tasks. In particular, our chemical tagging experiments and associated datasets could be useful in comparing different supervised disentanglement architectures, something that has so far been lacking in the machine learning community. We believe that our task of evaluating how well a supervised disentanglement neural network maps chemically identical stars to an identical latent is particularly useful for assessing supervised disentanglement. This is because it does not rely on training any secondary networks and gives a single value that is directly indicative of the level of disentanglement found in the latent. Our proposed novel supervised disentanglement architecture has shown good performance at chemical tagging-like pursuits, and disentanglement which suggests that it may be a competitive alternative to Fader disentanglement types of architectures.

\begin{acknowledgments}

DDM is supported by the STFC UCL Centre for Doctoral Training in Data Intensive Science (grant number ST/P006736/1). The authors also acknowledge the Flatiron institute for use of their HPC system which allowed this work to  be  performed.  The  authors  would  also  like  to  thank  Ioanna Manolopoulou for helpful discussions, Jo Bovy for maintaining the APOGEE python package and David W. Hogg for helping connect the co-authors.

\end{acknowledgments}

\pagebreak

\appendix

\section{Neural Network Training Details}\label{appendix:training}
We briefly review some implementation details useful for reproducing the results in this paper. We have made our repository open-source to aid in making our paper reproducible and encourage readers to refer to the code for additional details.

\textbf{Dataset processing:} We process the continuum-normalized spectra by first multiplying the spectra by 4 and then substracting by 3.5. This makes the spectra roughly occupy the [-1,1] range.

\textbf{Neural network training:} All quoted results use feedforward neural networks with self-normalized rectified units (SELU) activation functions \citep{SELU}. All results are obtained using the ADAM optimizer \citep{ADAM} with a learning rate of $10^{-5}$. In the following, $n_{\rm bins}=7751$ refers to the number of spectral bins used, $n_{\rm conditioned}=2 \rm \: or\: 3$ the number of parameters the encoder is conditioned on, and $n_{\rm z}=20$ the size of the autoencoder latent.

\textbf{FactorDis architecture:} Our FactorDis neural network has the following architecture (including input and output layers). Results can be reproduced with loss weighting term $\lambda=10^{-4}$

\begin{equation}
    \text{encoder dimensions} = \{n_{\rm bins} +n_{\rm conditioned}, 2048 , 512, 128, 32, n_{z}\}
\end{equation}

\begin{equation}
    \text{decoder dimensions} = \{n_{\rm z}+n_{\rm conditioned}, 512, 2048, 8192, n_{bins}\}
\end{equation}

\begin{equation}
    \text{discriminator dimensions} = \{n_{\rm bins} +n_{\rm conditioned}+n_{\rm z}, 4096, 1024, 512, 128, 32, 1\}
\end{equation}

\textbf{FaderDis architecture:} Our FaderDis neural network has the following architecture (including input and output layers). Results can be reproduced with loss weighting term $\lambda = 10^{-5}$.  When training the auxiliary network, each disentangled parameter was split into 10 discrete values creating 100 equal sized bins when disentangling two parameters and 1000 equal sized bins when disentangling three.

\begin{equation}
    \text{encoder dimensions} = \{n_{\rm bins} +n_{\rm conditioned}, 2048 , 512, 128, 32, n_{\rm z}\}
\end{equation}

\begin{equation}
    \text{decoder dimensions} = \{n_{\rm z}+n_{\rm conditioned}, 512, 2048, 8192,n_{\rm bins}\}
\end{equation}

\begin{equation}
    \text{auxilary dimensions} = \{n_{\rm z}+n_{\rm conditioned}, 512, 256, 10^{n_{\rm conditioned}}\}
\end{equation}

\textbf{Non-linear Chemical Estimation:} In Figure \ref{fig:interpretability}, neural networks, taking latents z as inputs, are used as non-linear estimators of abundances. A separate neural network with the following structure was trained for every chemical specie.
\begin{equation}
    \text{non-linear dimensions} = \{n_z,512,256,128,1\}
\end{equation}


\bibliography{references}
\bibliographystyle{aasjournal}




\end{document}